\documentclass[a4paper]{jpconf}
\usepackage{graphicx}
\begin{document}
\title{Central Compact Objects in Supernova Remnants}

\author{Andrea De Luca}

\address{INAF/IASF Milano, Via Bassini 15, I-20133 Milano, Italy}

\ead{deluca@iasf-milano.inaf.it}

\begin{abstract}
Central Compact Objects (CCOs) are a handful of sources located close to the geometrical center of young supernova remnants. 
They only show thermal-like, soft X-ray emission and have no counterparts at any other wavelength. While the first observed 
CCO turned out to be a very peculiar magnetar, discovery that three members of the family are weakly magnetised Isolated Neutron Stars (INSs) 
set the basis for an interpretation of the class. 
However, the phenomeology of CCOs  and their relationship with other classes of INSs, 
possibly ruled by supernova fall-back accretion, are still far from being well understood.
\end{abstract}

\section{Introduction}
A dozen of point-like, soft X-ray sources has been discovered over the last two decades, gently shining very close to the geometrical center of young (0.3--7 kyr) supernova remnants (SNR). Their spectra are thermal-like, usually well described by the sum of two blackbodies with high temperatures (0.2--0.5 keV) and very small emitting radii (ranging from 0.1 to a few km). Their emission is generally steady, with a luminosity of the order of $10^{33}$ erg s$^{-1}$, and pulsations are undetected in most cases. They have no counterparts at any wavelength. No associated structures of diffuse, non-thermal emission are seen. Such sources are clearly different from both standard rotation-powered pulsars and magnetars. They have been dubbed, as a class, ``Central Compact Objects'' (CCOs)\footnote{The name has been used for the first time by G. Pavlov, referring to the central source in the Cas A SNR \cite{2000ApJ...531L..53P}.} -- indeed, a designation suggesting our rather poor understanding of their nature and physics. 
Discovery of fast pulsations and measurement of a tiny, positive period derivative in three sources
set the basis of a framework for CCOs as 
young, Isolated Neutron Stars (INS) with a weak dipole field. 
In this short paper, I will focus on recent results and directions in CCO studies, including discovery
of the magnetar nature of the first observed CCO -- a possibly unique source.  I will show that
an explanation of the properties of these sources would be highly relevant for our overall 
understanding of NS production rate, physics of core-collapse, INS diversity and evolution.
Previous reviews on CCOs were given by 
\cite{2004IAUS..218..239P,2008AIPC..983..311D,2013ApJ...765...58G}.

\section{1E 161348--5055: a unique, slowly-rotating magnetar.}
1E 161348--5055 (1E 1613) in the 2 kyr-old SNR RCW103 has been the prototype for CCOs. 
It was the first candidate radio-quiet INS 
discovered in a SNR \cite{1980ApJ...239L.107T}.
However, more recent observations clearly  separated 1E 1613 from CCOs and 
settled the case for a truly unique phenomenology: 1E 1613 displays
a dramatic long-term variability with large outbursts and a puzzling periodicity at 6.67 hours,
together with a young age and lack of an optical/infrared counterpart \cite{2006Sci...313..814D,2008ApJ...682.1185D}. 
The nature of the source, possibly the first low-mass X-ray binary system observed inside a SNR, 
or a very peculiar isolated, young magnetar with an extremely slow spin period, possibly braked 
by interaction with a surrounding disk \cite{2006Sci...313..814D}
remained debated for a decade - even more
exotic pictures were proposed \cite{2007ApJ...666L..81L,2008ApJ...681..530P,2009A&A...506.1297B,2011MNRAS.418..170E,2013Ap&SS.346..105I,2015ApJ...799..233L}.

A partial answer to this puzzle  came only very recently.
Indeed, on 2016 June 22, the Burst Alert Telescope onboard {\em Swift} 
detected a magnetar-like, short X-ray burst from the direction of 1E 1613,
with a spectrum well described by a blackbody model ($kT\sim9$ keV)
and a luminosity of $\sim2\times10^{39}$ erg s$^{-1}$. 
A strong X-ray outburst was simultaneously observed from 1E 1613 with the {\em Swift} X-Ray Telescope (XRT), 
with the 0.5--10 keV flux a factor of 100 brighter than 
the quiescent level observed up to one month 
before \cite{2016ApJ...828L..13R,2016MNRAS.463.2394D}.
Follow-up observations performed 
with {\em Chandra} and {\em NuSTAR} unveiled (i) a dramatic change 
in the shape of the 6.67 hour modulation,  two broad peaks per period replacing the 
sinusoidal shape that had been observed since 2005  \cite{2016ApJ...828L..13R}; (ii) 
a non-thermal spectral component extending up to $\sim30$ keV (modulated up to $\sim20$ keV), 
never detected before, well described by a hard power law with $\Gamma\sim1.2$, 
superimposed to the thermal continuum dominating at lower energies \cite{2016ApJ...828L..13R}.
Based on a series of {\em Swift}/XRT observations, the overall energy emitted in the outburst
(impulsive plus persistent) was computed to be  $\sim2.6\times10^{42}$ ergs
\cite{2016ApJ...828L..13R}.
A likely counterpart in the near infrared was detected by \cite{2017ApJ...841...11T} with 
the {\em Hubble Space Telescope} 
with AB magnitudes of 26.3 and 24.2 in the F110W and F160W filters, respectively (roughly corresponding to the J and H bands). 
The same source was detected with the ESO/{\em VLT} 
with Ks=$20.9\pm0.1$ (De Luca et al., in preparation).
All results from such multiwavelength observations 
strongly point to a magnetar interpretation for 1E 1613 \cite{2016ApJ...828L..13R,2016MNRAS.463.2394D,2017ApJ...841...11T}.
Indeed, 1E 1613 stands out as a unique source because of its rotation period, 
slower by three orders of magnitude with respect to any other known magnetar candidate. 

Assuming that 1E 1613 was born with a spin period of a few hours 
is hardly conciliable with angular momentum conservation in the progenitor star's collapsing core and also
clashes with the commonly assumed dynamo mechanism for the generation of the gigantic 
magnetic field in magnetars, requiring a fast initial period ($\sim$millisecond) \cite{1993ApJ...408..194T}.
Disregarding such a possibility, 
one should explain how a fast-spinning NS could be slowed-down to a period of 6.67 hours in $\sim2$ kyr.  
Standard magneto-dipole braking, coupled to magneto-thermal evolution, cannot account for such a slow rotation
for any reasonable assumption on the newborn NS properties \cite{2016ApJ...828L..13R}.
Braking of a strongly-magnetized NS by material/magnetic interaction 
with a low-mass companion star was discussed by \cite{2008ApJ...681..530P}. However, near-infrared observations rule out
any possible companion with a mass larger than a M8 dwarf \cite{2017ApJ...841...11T} -- it seems unlikely that a binary system with such an
extreme mass ratio survive a supernova explosion\footnote{The same argument, together with the difficulty in explaining the 
dramatic long-term variability in pulse shape and pulsed fraction, argues against an interpretation of the 6.67 hr cycle
as the orbital period in a binary.}. An alternative possibility is to invoke the role of 
supernova fallback material \cite{1971ApJ...163..221C,1989ApJ...346..847C}.
The NS could have been braked by propeller interaction with a long-lived residual disk. Such a possibility had been considered 
by \cite{2006Sci...313..814D,2007ApJ...666L..81L} and, more recently, by \cite{2016ApJ...833..265T,2017MNRAS.464L..65H}. Although
the formation and stability of fallback disks surrounding newborn NSs has been questioned by \cite{2014ApJ...781..119P},
it was shown that a very low-mass disk ($\sim10^{-9}$ M$_{\odot}$) could  brake the NS to the observed spin rate
in $\sim1-3$ kyr, provided it is endowed by a dipole magnetic field of $5\times10^{15}$ G \cite{2017MNRAS.464L..65H},
the largest in the magnetar family.
As a further possibility, the NS could have been slowed-down in an earlier phase -- propeller interaction starting at the onset 
of fallback accretion (with no need for formation and long-term survival of a disk). Such a picture was mentioned by \cite{2016ApJ...828L..13R,2015PASA...32...18P}, 
but no detailed calculations have been presented so far. 

\section{CCO pulsars: neutron stars with a weak dipole magnetic field.}
A key discovery 
has been the detection of fast pulsations from three objects of the class: 
1E 1207.4-5209 in G296.5+10.0 ($P\sim424$ ms, \cite{2000ApJ...540L..25Z}), CXOU J185238.6+004020 in Kes 79 
($P\sim105$ ms, \cite{2005ApJ...627..390G})
and RX J0852.0-4622 in Puppis A ($P\sim 112$ ms, \cite{2009ApJ...695L..35G}). This proved with no doubt that these sources are 
NSs. Even more enlightening has been the measurement of their period derivatives \cite{2013ApJ...765...58G,2010ApJ...709..436H},
which turned out to be very small 
(indeed, very long observation campaigns were required). Implications are
extremely interesting: (i) the dipole magnetic field inferred from standard magneto-dipole braking $B_{dip}=3.2\times10^{19} (P \dot{P})^{1/2}$
 is $10^{10}-10^{11}$ G, 
remarkably smaller than the one of the bulk of the rotation-powered pulsar population; (ii) the characteristic age ($\tau_C=P/2\dot{P}$) 
is 4--5 orders of magnitude larger than the age of the host SNR (indeed, the CCO birth period is likely very close to the currently observed one); 
(iii) the spin-down energy loss is more than 10 times smaller than the X-ray luminosity. 
However, several properties of the CCO pulsars do not fit easily in the picture of weakly magnetised INSs. 

CXOU J185238.6+004020 displays an extremely high modulation (pulsed fraction $\sim64\%$, with very little dependence on energy) 
of its thermal emission, consisting of the sum of two hot blackbodies with tiny emitting areas 
 \cite{2010ApJ...709..436H}. Reproducing the observed pulse shape and fraction with thermal emission models
is challenging \cite{2012ApJ...748..148S} 
and points to a peculiar thermal map, with a high temperature contrast between a small emitting region (a hot spot surrounded by a warmer region)
and the remaining, cooler and unobservable surface of the star. A highly elongated shape in the longitudinal direction (with respect to the spin axis) 
is required for the emitting region; a more conventional, polar cap geometry requires the radiation to be highly beamed \cite{2014ApJ...790...94B}. 

RX J0852.0-4622 too has a thermal continuum well described by two blackbodies. 
Its pulse shape is sinusoidal,
with a very  abrupt, $180^{\circ}$ phase reversal at $\sim1.2$ keV - where the hot and warm blackbodies
switch dominance \cite{2010ApJ...724.1316G}. This is consistent with an anti-podal hot spot model, with two small  emitting regions 
with different size and temperature located at opposite sides of the NS \cite{2010ApJ...724.1316G}. A spectral feature is also seen
in the low-energy portion of the spectrum. It can be described either as an emission line at $\sim0.75$ keV, or as a couple
of absorption lines at $\sim0.46$ keV and $\sim0.92$ keV. Significant variability in the feature is seen between two observations performed 
in 2001 and 2009: assuming the emission line model, the central energy  decreases from $\sim0.79$ to $\sim0.71$ keV 
\cite{2013ApJ...765...58G,2012MNRAS.421L..72D}. No further variability is seen  after 2009 \cite{2013ApJ...765...58G}.

1E 1207.4-5209 displays multiple absorption features at harmonically spaced energies (0.7, 1.4, 2.1 and possibly 2.8 keV) superimposed
to a thermal spectrum \cite{2002ApJ...574L..61S,2002ApJ...581.1280M,2003Natur.423..725B,2004A&A...418..625D}.
After a long debate (see \cite{2008AIPC..983..311D}), measurement of $P$ and $\dot{P}$ points to an interpretation of the features as due to electron cyclotron
scattering close to the star's surface, the feature at 0.7 keV being the fundamental. The magnetic field strength  in the region where the lines
are formed would be $\sim8\times10^{10}$ G (accounting for a gravitational redshift $z=0.3$), in broad agreement with
the value inferred by magneto-dipole braking ($\sim9.8\times10^{10}$ G). The relative strength of the harmonics was 
explained by taking into account resonances in the photospheric free-free opacity in the presence of the
magnetic field \cite{2010ApJ...714..630S,2010A&A...518A..24P,2012ApJ...751...15S}. The X-ray pulsation is dominated
by the complex modulation of the spectral features as a function of the rotational phase \cite{2004A&A...418..625D},
the continuum, well described by the sum of two blackbody curves, 
being almost unpulsed.
No simple constraints on the geometry of the thermally emitting regions could be set.

\section{Other CCOs:  homogeneous group or mixed bag ?}

The family of CCOs includes about ten more sources\footnote{see {\tt www.iasf-milano.inaf.it/\~{}deluca/cco/main.htm} for an updated list}. 
Thermal emission properties 
are pretty homogeneous, with no pulsations (rather deep upper limits have been set in a few cases), 
nor long-term variability (although multi-epoch coverage is limited in most cases). It seems reasonable to assume that
such sources are INS similar to the three CCO pulsars, possibly with even smaller dipole fields.

The most famous CCO
was discovered close to the center 
of the very young ($\sim350$ yr) Cas A SNR in the Chandra ``first light'' image \cite{1999IAUC.7246....1T}. 
An early claim of a magnetar nature for this CCO,
based on detection of the possible infrared echo of a giant flare occurred around 1950 \cite{2005Sci...308.1604K} 
was later retracted \cite{2008ApJ...678..287K,2008ApJ...685..976D}.
Multi-epoch observations with Chandra point to  long-term evolution of the spectral shape
and luminosity of this source, consistent with a $\sim4\%$ temperature decrease in about 10 yr \cite{2010ApJ...719L.167H}.
Such a direct observation of the NS cooling would have very important implications for our understanding
of the properties of ultra-dense matter in the NS interior 
\cite{2011MNRAS.412L.108S,2011MNRAS.411.1977Y,2011PhRvL.106h1101P,2011ApJ...735L..29Y,2013ApJ...765....1N}. 
However, more recent analysis of Chandra data focused 
on possible systematics affecting the measurement (e.g. instrument calibration issues)
and concluded that the reality and rate of the possible
temperature decrease are uncertain \cite{2013ApJ...777...22E,2013ApJ...779..186P}. 
 
It was shown \cite{2009Natur.462...71H} that a carbon atmosphere NS model with low magnetic field provides a good description 
of the spectrum of this source and implies an emitting region consistent with the entire NS surface. 
This possibly solves the puzzling lack of pulsations 
coupled to the large temperature anisotropy (with uncomfortably small emitting areas) as derived from other spectral models
(both blackbody curves and weakly-magnetised NS hydrogen atmosphere models),
pointing to the picture of a weakly magnetized NS, with uniform emission from its surface and with active nuclear burning in its surface layers.

The same picture was proposed for other CCOs, as soon as high-quality X-ray spectra became available: the source
in  G353.6-0.7 \cite{2013A&A...556A..41K,2015A&A...573A..53K} (possibly the hottest known INS), as well as
the source in G15.9+0.2 \cite{2016A&A...592L..12K}. It was stressed that the carbon atmosphere models makes it possible
to reconcile thermal luminosities
with the current distance estimates and also  to constrain the equation of state of the NS 
\cite{2015A&A...573A..53K,2015MNRAS.454.2668O}. As a matter of fact, however, 
such models cannot explain the phenomenology of all CCOs. For instance, the carbon atmosphere model would describe
the emission of the CCO pulsar in Kes 79 as coming from the whole surface of the NS, but this is clearly 
at odds with the observed, very high pulsed fraction \cite{2014ApJ...790...94B}. Indeed, the current upper limits on 
the CCO pulsations are not stringent enough to rule out highly anisotropic surface temperature distributions
\cite{2009ApJ...703..910P,2017A&A...600A..43S} that would be consistent with different atmosphere chemical composition as 
well as with different properties of the NSs.


\section{CCOs as a ``class''.}
\label{ccoclass}
The picture of CCOs as weakly-magnetised, young NSs has to face two issues. 

First issue: the highly-anisotropic,
high contrast-ratio temperature surface distribution cannot be easily explained in this framework. The hot spots cannot be heated by rotation-powered
particle bombardment (ruled out by the spin-down energetics); they cannot be accretion-powered (ruled out by 
timing \cite{2010ApJ...709..436H} as well as by deep optical limits to any companion star, e.g. \cite{2011A&A...525A.106D}). 
Hot spots could be due to localized crustal heating by 
magnetic field decay. This would require a magnetic field with strong  non-dipolar components ($10^{14}-10^{15}$ G 
to account for the observed luminosity \cite{2010ApJ...709..436H}), coupled to a factor $10^4$ less intense dipolar component.
As an alternative possibility,  hot spots could be
powered by residual heat. This would require highly anisotropic heat transfer from the NS interior, pointing to a strong
magnetic field in the crust. For instance, a toroidal field of order $10^{14}$ G could screen large part of the surface
around the equator, channelling the heat flux towards the poles \cite{2012ApJ...748..148S,2006A&A...451.1009P}. 
A poloidal field as low as $10^{10}-10^{11}$ G
could be enough to make the magnetic field configuration stable
\cite{2009MNRAS.397..763B}. 

Second issue: the region in the $P-\dot{P}$ parameter space where CCO pulsars are located 
is highly underpopulated, at odds with expectations \cite{2010PNAS..107.7147K}.
Such region is well above the ``death line'' for radio pulsars. Indeed, rotation-powered radio emission 
is seen from several sources with similar spin parameters. Moreover, CCOs are a common outcome of core-collapse 
supernovae -- they are found in SNR as frequently as other classes of NSs 
(more frequently than magnetars, see e.g. \cite{2010ApJ...710..941H}). 
Frequent formation, coupled with a very slow evolution of the spin period would 
lead to a large number of low $B$-field sources above the death line. 
As a possible explanation of this puzzle, CCOs could be intrinsically radio-quiet
and become essentially invisible after their host SNR fade away.
Alternatively, radio luminosity in rotation-powered emission could depend on spin-down power, as suggested by radio pulsar population
synthesis (e.g. \cite{2006ApJ...643..332F}). As a further possibility, CCOs could evolve 
and move to a different region of the $P-\dot{P}$ diagram. This would ease a further, related issue: the sum of the inferred 
birth rates for different classes of NSs exceeds the Galactic core-collapse supernova rate (see e.g. \cite{2008MNRAS.391.2009K},
although these authors did not discuss the case of CCOs)
-- this points to evolutionary links among different NS families. 

Different frameworks for CCOs as young INSs with weak dipole field have been discussed. 
A first possible scenario \cite{2007ApJ...664L..35G,2007ApJ...665.1304H} links the origin of the weak field  
to a slow rotation of the collapsing progenitor star core, which would result in an inefficient dynamo mechanism \cite{2006A&A...451.1049B}.
Such a picture 
does not seem very appealing: the birth period for CCOs is not ``long'', according to population synthesis models for  radio pulsars 
(e.g. \cite{2006ApJ...643..332F} point to a wide distribution of birth periods with a mean  of 300 ms and a dispersion of 150 ms). 
Moreover, the origin of the peculiar thermal anisotropy would remain unexplained. 
As a second possibility, CCOs could be 
quiescent magnetars, with an extremely weak dipole field, but with a strong crustal magnetic field, emerging 
in local ``sunspot'' structures. 
Such a scenario, however, seems rather unlikely 
because of the general lack of variability in CCOs, at odds with the ubiquitous variability seen in magnetars 
(but we cannot exclude the picture to be correct for a fraction of the sample). 
A third picture is known as the ``buried field'' scenario:
prompt accretion of supernova fallback material \cite{1971ApJ...163..221C}
could bury the magnetic field of the newborn NS beneath its surface; the field could then re-emerge 
by diffusion on a time 
scale depending on the amount of accreted matter \cite{1989ApJ...346..847C,1995ApJ...440L..77M,1999A&A...345..847G}.
Recent models suggest that spherical accretion of $\sim10^{-4}-10^{-2}$ M$_{\odot}$ is required to screen typical fields
in the $10^{12}-10^{14}$ G range for $\sim10^3-10^5$ yr
\cite{2011MNRAS.414.2567H,2012MNRAS.425.2487V,2013ApJ...770..106B,2016MNRAS.456.3813T,2016MNRAS.462.3646B,2016MNRAS.462.3689I}.
According to these studies, (i) the strong, hidden crustal field could explain peculiarities in the thermal map;
(ii) a low dipole field could be common in young NSs ($\leq$ few kyr);  (iii) CCOs could turn in radio pulsars 
at an older age (see also \cite{2016MNRAS.457.1180R}).
It is  difficult to test the buried field scenario on CCOs by X-ray timing: evidence for a re-emerging magnetic field could 
come from a measurement of their braking index $n=2-P\ddot{P}/\dot{P}^2$, expected to be equal to 3 
in case of pure magneto-dipole braking for a constant magnetic field, and smaller than 3
in case of a growing dipole field. However, the tiny $\dot{P}$ of CCOs make any measurement
of their $n$ extremely challenging. Some supporting evidence for this picture is provided by timing investigation of young ($\tau_c\sim10^3-10^4$ yr) rotation-powered pulsars,
yielding low values for the braking index that can be interpreted as hint of  a growing magnetic field \cite{2015MNRAS.452..845H,2016ApJ...827L..39M};
indirect support is also provided by the absence of evidence for weakly magnetised ($\leq10^{11}$ G) NSs in High-mass X-ray binaries,
consistent with field re-emergence in these sources on a time scale of $\sim10^4$ yr \cite{2012ASPC..466..191P}.

A different way to attack the problem is to search for CCO descendants. Such investigations assume that at least some CCO may be 
a radio pulsar; CCOs would be much younger than ``standard'' pulsars in a given $P-\dot{P}$ parameter space region and thus 
they should be distinguishable because of a high thermal luminosity. A search for CCO descendants was performed 
among ``mildly-recycled'' pulsars (NSs that have accreted gas from a companion star in a binary system
for a short time before a second supernova explosion halted their spin-up) -- a class of pulsars proposed to explain the sparse  
population of sources in the same $P-\dot{P}$ region where  CCOs have been recently located. However, none was found \cite{2013ApJ...773..141G}.
A second attempt was carried out, considering seemingly old radio pulsars spatially coincident with known SNRs,
 with negative results \cite{2014ApJ...792L..36B}. A third investigation did not identify
 any plausible CCO among radio pulsars with weak field ($B<10^{11}$ G), but energetics larger than mildly recycled pulsars
 \cite{2015ApJ...808..130L}. These studies suggest that CCO descendants, if not radio-quiet, should hide
 among pulsars with larger B-field and energetics. Interestingly, the peculiar neutron star dubbed {\em Calvera} \cite{2008ApJ...672.1137R} has been proposed as a 
 possible example of an evolved CCO, whose SNR is no more detectable in X-rays. 
 It is a radio-quiet pulsar, with $P\sim59$ ms \cite{2011MNRAS.410.2428Z} and $\dot{E}_{rot}\sim6\times10^{35}$ erg s$^{-1}$
 \cite{2013ApJ...778..120H}, only showing  soft X-ray, thermal-like emission in spite of its high spin-down luminosity.
 Calvera lies in the $P-\dot{P}$ plane along the path of growing $B$-field between CCOs and young radio pulsars; however, a lack of any constraint on its distance (hence on its
 luminosity) as well as on its true age does not allow us to draw a firm conclusion \cite{2015ApJ...812...61H}.

\section{Conclusions.}
\label{conclusions}
CCOs, a handful of candidate young INSs  with elusive properties, 
proved to include extraordinarily interesting objects. 
Explanation of their phenomenology challenges standard models and points to
a very complex scenario, in which accretion of supernova fallback material in different 
regimes, coupled to a variety 
of initial conditions for the magnetic field strength and configuration, 
plays an important role in shaping the properties of newborn NSs. This has very important implications
towards a physical understanding of the different phenomenological classes of INSs
and of their evolution. There is still a large space for discoveries in CCO studies, especially in view 
of forecoming observing facilities such as eROSITA, ATHENA, SKA. Deep searches for periodicity and long-term flux monitoring in the X-rays
will allow one to assess the nature of non-pulsating CCOs - proving them to be similar to the three CCO pulsars,
or unveiling any (peculiar) magnetar among them - as a matter of fact, the spectrum of a quiescent magnetar and of a CCO are almost undistinguishable. 
Large surveys will possibly allow one to identify new candidate CCOs and/or
candidate CCO descendants. (Targeted) radio pulsar searches will unveil if CCOs are intrinsically radio-quiet, or simply radio-faint. Using already available data,
modeling the phase-resolved behaviour of CCO pulsar will allow one to constrain the physics of the spectral features, the surface thermal map and magnetic field topology, 
as well as to test expectations of the buried field scenario. 


\subsection{Acknowledgments}
I warmly thank the SOC and the LOC for organizing a very interesting meeting and for the invitation to speak.
I thank P. Esposito for a critical reading of the manuscript and G.F. Bignami (who left us prematurely on May 24$^{th}$, 2017), A. Borghese, P.A. Caraveo, F. Coti Zelati, 
 G.L. Israel, S. Mereghetti, N. Rea, A. Tiengo for many useful discussions.

\end{document}